
\input epsf
\newbox\leftpage \newdimen\fullhsize \newdimen\hstitle \newdimen\hsbody
\tolerance=1000\hfuzz=2pt
\catcode`\@=11 
\magnification=\magstephalf\baselineskip=16pt plus 2pt minus 1pt
\hsbody=\hsize \hstitle=\hsize 
%

%
%

\def\draftmode{\message{ DRAFTMODE }\def\draftdate{{\rm preliminary draft:
\number\month/\number\day/\number\yearltd\ \ \hourmin}}%
\headline={\hfil\draftdate}\writelabels\baselineskip=20pt plus 2pt minus 2pt
 {\count255=\time\divide\count255 by 60 \xdef\hourmin{\number\count255}
  \multiply\count255 by-60\advance\count255 by\time
  \xdef\hourmin{\hourmin:\ifnum\count255<10 0\fi\the\count255}}}
\def\nolabels{\def\wrlabeL##1{}\def\eqlabeL##1{}\def\reflabeL##1{}}
\def\writelabels{\def\wrlabeL##1{\leavevmode\vadjust{\rlap{\smash%
{\line{{\escapechar=` \hfill\rlap{\sevenrm\hskip.03in\string##1}}}}}}}%
\def\eqlabeL##1{{\escapechar-1\rlap{\sevenrm\hskip.05in\string##1}}}%
\def\reflabeL##1{\noexpand\llap{\noexpand\sevenrm\string\string\string##1}}}
\nolabels
%
\global\newcount\secno \global\secno=0
\global\newcount\meqno \global\meqno=1
\def\newsec#1{\global\advance\secno by1\message{(\the\secno. #1)}
\global\subsecno=0\eqnres@t\noindent{\bf\the\secno. #1}
\writetoca{{\secsym} {#1}}\par\nobreak\medskip\nobreak}
\def\eqnres@t{\xdef\secsym{\the\secno.}\global\meqno=1\bigbreak\bigskip}
\def\sequentialequations{\def\eqnres@t{\bigbreak}}\xdef\secsym{}
\global\newcount\subsecno \global\subsecno=0
\def\subsec#1{\global\advance\subsecno by1\message{(\secsym\the\subsecno. #1)}
\ifnum\lastpenalty>9000\else\bigbreak\fi
\noindent{\it\secsym\the\subsecno. #1}\writetoca{\string\quad 
{\secsym\the\subsecno.} {#1}}\par\nobreak\medskip\nobreak}
\def\appendix#1#2{\global\meqno=1\global\subsecno=0\xdef\secsym{\hbox{#1.}}
\bigbreak\bigskip\noindent{\bf Appendix #1. #2}\message{(#1. #2)}
\writetoca{Appendix {#1.} {#2}}\par\nobreak\medskip\nobreak}
%
%
\def\eqnn#1{\xdef #1{(\secsym\the\meqno)}\writedef{#1\leftbracket#1}%
\global\advance\meqno by1\wrlabeL#1}
\def\eqna#1{\xdef #1##1{\hbox{$(\secsym\the\meqno##1)$}}
\writedef{#1\numbersign1\leftbracket#1{\numbersign1}}%
\global\advance\meqno by1\wrlabeL{#1$\{\}$}}
\def\eqn#1#2{\xdef #1{(\secsym\the\meqno)}\writedef{#1\leftbracket#1}%
\global\advance\meqno by1$$#2\eqno#1\eqlabeL#1$$}
%
\newskip\footskip\footskip14pt plus 1pt minus 1pt 
\def\footnotefont{\ninepoint}\def\f@t#1{\footnotefont #1\@foot}
\def\f@@t{\baselineskip\footskip\bgroup\footnotefont\aftergroup\@foot\let\next}
\setbox\strutbox=\hbox{\vrule height9.5pt depth4.5pt width0pt}
\global\newcount\ftno \global\ftno=0
\def\foot{\global\advance\ftno by1\footnote{$^{\the\ftno}$}}
%
\newwrite\ftfile   
\def\footend{\def\foot{\global\advance\ftno by1\chardef\wfile=\ftfile
$^{\the\ftno}$\ifnum\ftno=1\immediate\openout\ftfile=foots.tmp\fi%
\immediate\write\ftfile{\noexpand\smallskip%
\noexpand\item{f\the\ftno:\ }\pctsign}\findarg}%
\def\footatend{\vfill\eject\immediate\closeout\ftfile{\parindent=20pt
\centerline{\bf Footnotes}\nobreak\bigskip\input foots.tmp }}}
\def\footatend{}
%
%
\global\newcount\refno \global\refno=1
\newwrite\rfile
\def\ref{[\the\refno]\nref}
\def\nref#1{\xdef#1{[\the\refno]}\writedef{#1\leftbracket#1}%
\ifnum\refno=1\immediate\openout\rfile=refs.tmp\fi
\global\advance\refno by1\chardef\wfile=\rfile\immediate
\write\rfile{\noexpand\item{#1\ }\reflabeL{#1\hskip.31in}\pctsign}\findarg}
\def\findarg#1#{\begingroup\obeylines\newlinechar=`\^^M\pass@rg}
{\obeylines\gdef\pass@rg#1{\writ@line\relax #1^^M\hbox{}^^M}%
\gdef\writ@line#1^^M{\expandafter\toks0\expandafter{\striprel@x #1}%
\edef\next{\the\toks0}\ifx\next\em@rk\let\next=\endgroup\else\ifx\next\empty%
\else\immediate\write\wfile{\the\toks0}\fi\let\next=\writ@line\fi\next\relax}}
\def\striprel@x#1{} \def\em@rk{\hbox{}} 
\def\lref{\begingroup\obeylines\lr@f}
\def\lr@f#1#2{\gdef#1{\ref#1{#2}}\endgroup\unskip}
\def\semi{\hfil\break}
\def\addref#1{\immediate\write\rfile{\noexpand\item{}#1}} 
\def\startrefs#1{\immediate\openout\rfile=refs.tmp\refno=#1}
\def\xref{\expandafter\xr@f}\def\xr@f[#1]{#1}
\def\refs#1{\count255=1[\r@fs #1{\hbox{}}]}
\def\r@fs#1{\ifx\und@fined#1\message{reflabel \string#1 is undefined.}%
\nref#1{need to supply reference \string#1.}\fi%
\vphantom{\hphantom{#1}}\edef\next{#1}\ifx\next\em@rk\def\next{}%
\else\ifx\next#1\ifodd\count255\relax\xref#1\count255=0\fi%
\else#1\count255=1\fi\let\next=\r@fs\fi\next}
%

%
\newwrite\ffile\global\newcount\figno \global\figno=1
\def\fig{figure~\the\figno\nfig}
\def\nfig#1{\xdef#1{figure~\the\figno}%
\writedef{#1\leftbracket figure\noexpand~\the\figno}%
\ifnum\figno=1\immediate\openout\ffile=figs.tmp\fi\chardef\wfile=\ffile%
\immediate\write\ffile{\noexpand\medskip\noexpand\item{Fig.\ \the\figno. }
\reflabeL{#1\hskip.55in}\pctsign}\global\advance\figno by1\findarg}
\def\xfig{\expandafter\xf@g}\def\xf@g figure\penalty\@M\ {}
\def\figs#1{figs.~\f@gs #1{\hbox{}}}
\def\f@gs#1{\edef\next{#1}\ifx\next\em@rk\def\next{}\else
\ifx\next#1\xfig #1\else#1\fi\let\next=\f@gs\fi\next}
\newwrite\lfile
{\escapechar-1\xdef\pctsign{\string\%}\xdef\leftbracket{\string\{}
\xdef\rightbracket{\string\}}\xdef\numbersign{\string\#}}

\def\writestop{\def\writestoppt{\immediate\write\lfile{\string\pageno%
\the\pageno\string\startrefs\leftbracket\the\refno\rightbracket%
\string\def\string\secsym\leftbracket\secsym\rightbracket%
\string\secno\the\secno\string\meqno\the\meqno}\immediate\closeout\lfile}}
\def\writestoppt{}\def\writedef#1{}
\def\seclab#1{\xdef #1{\the\secno}\writedef{#1\leftbracket#1}\wrlabeL{#1=#1}}
\def\subseclab#1{\xdef #1{\secsym\the\subsecno}%
\writedef{#1\leftbracket#1}\wrlabeL{#1=#1}}
\newwrite\tfile \def\writetoca#1{}
\def\leaderfill{\leaders\hbox to 1em{\hss.\hss}\hfill}
\def\writetoc{\immediate\openout\tfile=toc.tmp 
   \def\writetoca##1{{\edef\next{\write\tfile{\noindent ##1 
   \string\leaderfill {\noexpand\number\pageno} \par}}\next}}}

%
\catcode`\@=12 
%
\edef\tfontsize{\ifx\answ\bigans scaled\magstep3\else scaled\magstep4\fi}
 \tfontsize  \tfontsize
 \tfontsize \font\titlei=cmmi10 \tfontsize
\font\titleis=cmmi7 \tfontsize \font\titleiss=cmmi5 \tfontsize
\font\titlesy=cmsy10 \tfontsize \font\titlesys=cmsy7 \tfontsize
\font\titlesyss=cmsy5 \tfontsize  \tfontsize
\skewchar\titlei='177 \skewchar\titleis='177 \skewchar\titleiss='177
\skewchar\titlesy='60 \skewchar\titlesys='60 \skewchar\titlesyss='60
 \ifx\answ\bigans\else scaled\magstep1\fi
\ifx\answ\bigans\else

 \font\absi=cmmi10 scaled\magstep1
\font\absis=cmmi7 scaled\magstep1 \font\absiss=cmmi5 scaled\magstep1
\font\abssy=cmsy10 scaled\magstep1 \font\abssys=cmsy7 scaled\magstep1
\font\abssyss=cmsy5 scaled\magstep1 
\skewchar\absi='177 \skewchar\absis='177 \skewchar\absiss='177
\skewchar\abssy='60 \skewchar\abssys='60 \skewchar\abssyss='60
\fi
\font\ninerm=cmr9 \font\sixrm=cmr6 \font\ninei=cmmi9 \font\sixi=cmmi6 
\font\ninesy=cmsy9 \font\sixsy=cmsy6 \font\ninebf=cmbx9 
\font\nineit=cmti9 \font\ninesl=cmsl9 \skewchar\ninei='177
\skewchar\sixi='177 \skewchar\ninesy='60 \skewchar\sixsy='60 
\def\ninepoint{\def\rm{\fam0\ninerm}
\textfont0=\ninerm \scriptfont0=\sixrm \scriptscriptfont0=\fiverm
\textfont1=\ninei \scriptfont1=\sixi \scriptscriptfont1=\fivei
\textfont2=\ninesy \scriptfont2=\sixsy \scriptscriptfont2=\fivesy
\textfont\itfam=\ninei \def\it{\fam\itfam\nineit}\def\sl{\fam\slfam\ninesl}%
\textfont\bffam=\ninebf \def\bf{\fam\bffam\ninebf}\rm} 
\def\noblackbox{\overfullrule=0pt}

\def\etal{{\it et al}}

\def\frac#1#2{{{#1}\over {#2}}}
\def\half{\hbox{${1\over 2}$}}

\def\GeV{{\rm GeV}}

\def\MS{\hbox{$\overline{\rm MS}$}}

\catcode`@=11 
\def\slash#1{\mathord{\mathpalette\c@ncel#1}}
 \def\c@ncel#1#2{\ooalign{$\hfil#1\mkern1mu/\hfil$\crcr$#1#2$}}
\def\lsim{\mathrel{\mathpalette\@versim<}}
\def\gsim{\mathrel{\mathpalette\@versim>}}
 \def\@versim#1#2{\lower0.2ex\vbox{\baselineskip\z@skip\lineskip\z@skip
       \lineskiplimit\z@\ialign{$\m@th#1\hfil##$\crcr#2\crcr\sim\crcr}}}
\catcode`@=12 

\def\PR{{Physical Review~}}

\def\NP{{Nuclear Physics~}}

\def\PL{{Physics Letters~}}

\def\AP{{Ann~Phys~}}

\def\ZP{{Zeit~Phys~}}

\def\JHEP{{Jour~High~Energy~Phys~}}
\def\EPJ{{Euro~Phys~Jour~}}
\def\vol#1{{#1}}\def\vyp#1#2#3{\vol{#1} (#2) #3}
\def\footnoterule{\kern-3pt \hrule width \hsize \kern 2.6pt}
\def\pmb#1{\setbox0=\hbox{#1}
 \kern.05em\copy0\kern-\wd0 \kern-.025em\raise.0433em\box0 }
\font\fourteenbf=cmbx12 scaled\magstep1
\font\fourteenit=cmti12 scaled\magstep1
\parindent 0pt
\parskip 8pt
\noblackbox
\pageno=0\nopagenumbers\tolerance=10000\hfuzz=5pt
\baselineskip 12pt
\line{\hfill Edinburgh 99/21}
\line{\hfill DAMTP-1999-170}
\line{\hfill{\tt hep-ph/9912445}}
\vskip 12pt
\centerline{{\fourteenbf   THE CHALLENGE OF SMALL }{\fourteenit\pmb{x}}}
\vskip 18pt\centerline{R D  Ball\footnote{$^*$}{\footnotefont
Royal Society University Research Fellow}}
\vskip 6pt
\centerline{Department of Physics and Astronomy}
\centerline{University of Edinburgh, EH9 3JZ, Scotland{$^{\dag}$}}
\vskip 12pt
\centerline {P V ~Landshoff}
\vskip 6pt
\centerline {DAMTP, Centre for Mathematical Sciences}
\centerline {Cambridge, CB3 0AW, England\footnote{$^{\dag}$}{\footnotefont
email addresses: rdb@th.ph.ed.ac.uk \ \ pvl@damtp.cam.ac.uk}}
\vskip 50pt
\centerline{\bf Abstract}
\midinsert\leftskip 15truemm\rightskip 15truemm
{\narrower\baselineskip 11pt
\medskip\noindent
We review the current understanding of the behaviour of inclusive
cross sections at small $x$ and large $Q^2$ in terms of Altarelli-Parisi
evolution, the BFKL equation, and Regge theory, asking in particular
to what extent they are mutually consistent. 
\vskip 10mm
\centerline{\sl This report is a summary of various discussions 
at the} 
\centerline{\sl Durham phenomenology workshop, September 1999}
\smallskip}
\endinsert
\vfill
\line{December 1999\hfill}
\eject \footline={\hss\tenrm\folio\hss}

{\bf Introduction} 

A striking discovery at HERA has been the rapid rise with $1/x$ of the
proton structure $F_2$ at small $x$.  If one 
fits this rise to an effective power $x^{-\lambda(Q^2)}$
then, even at quite small values of $Q^2$, $\lambda (Q^2)$ is found to
be significantly greater than the value just less than 0.1
associated with soft pomeron exchange
that is familiar in purely hadronic collisions \ref\sigtot{
A Donnachie and P V Landshoff, \PL\vyp{B296}{1992}{227}
}. 
Moreover, $\lambda(Q^2)$ increases rapidly with~$Q^2$. 
Similarly, and perhaps equally importantly, the size of the scaling 
violations is seen to increase dramatically as we go to smaller $x$ 
(see figure~1).

\midinsert
\centerline{\epsfxsize=0.5\hsize\epsfbox{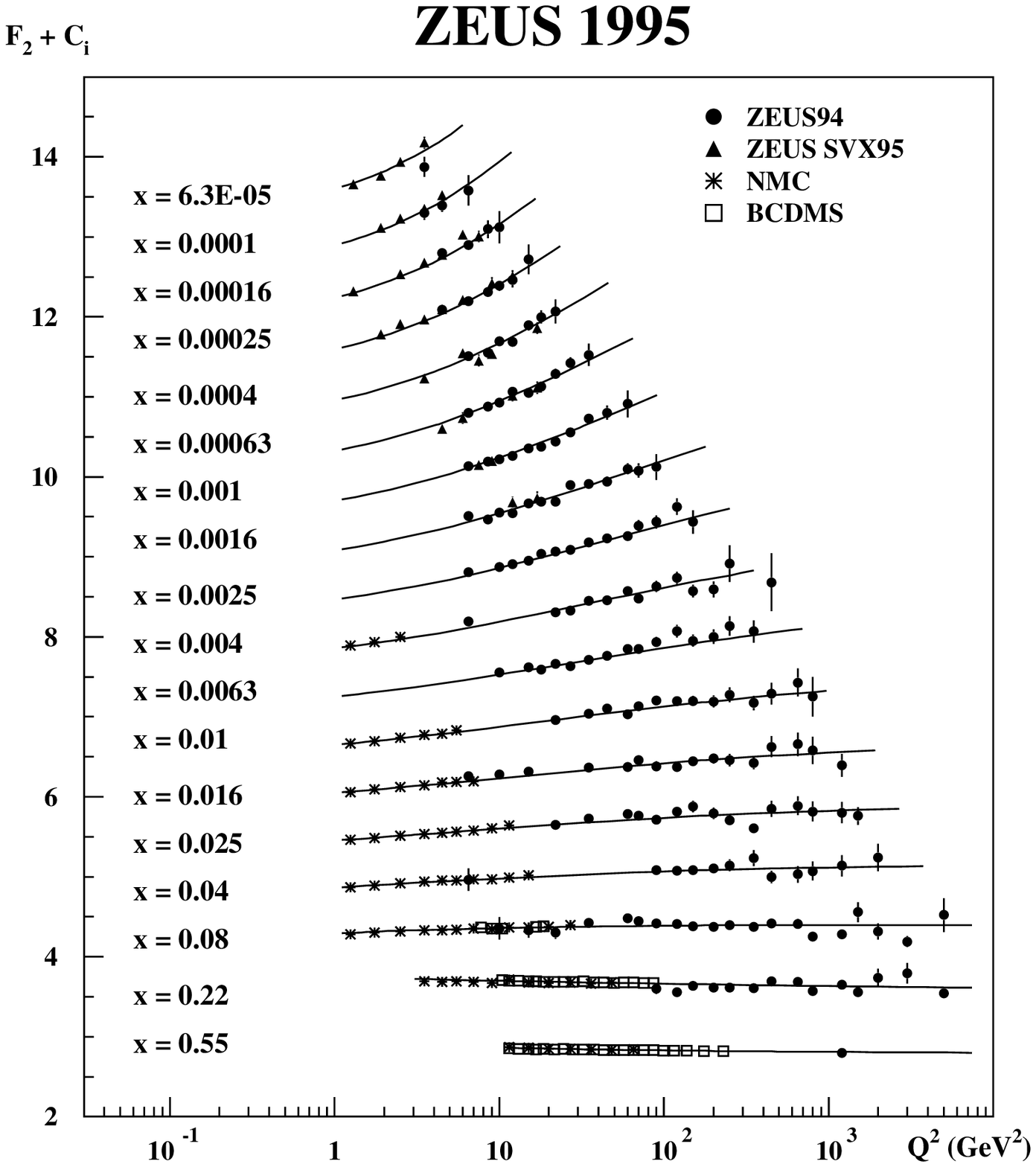}}
\centerline{\epsfxsize=0.5\hsize\epsfbox{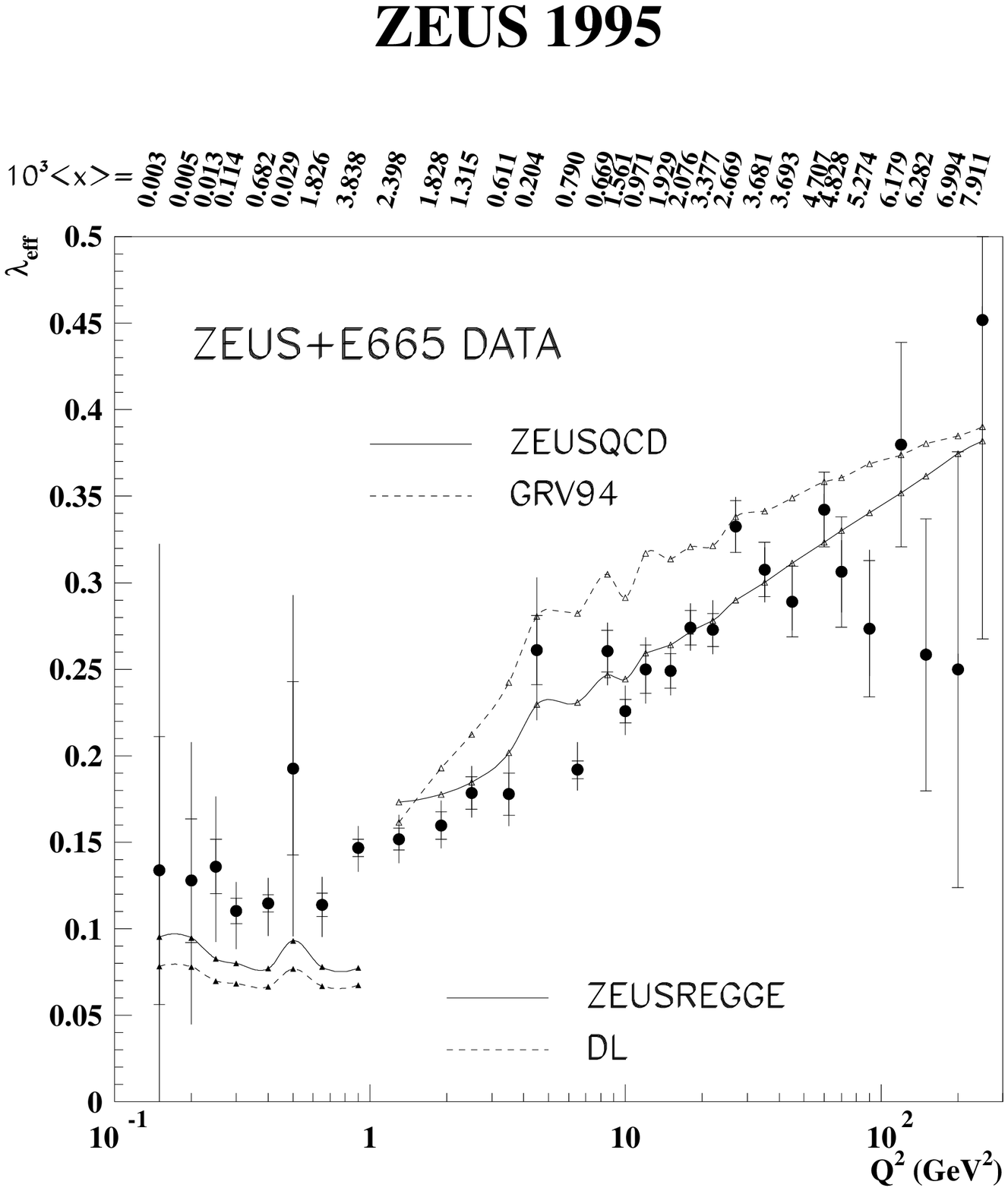}}
{\parindent=1cm \narrower\noindent
Figure 1: a) Measurements of $F_2$ by ZEUS \ref\ZEUS{ZEUS Collaboration,
\EPJ\vyp{C7}{1999}{609}}. The curves show a NLO
perturbative fit, with scaling violations as predicted by perturbative
QCD. b) $\lambda(Q^2)$ extracted from ZEUS and E665 data
on $F_2(x,Q^2)$ \ZEUS. The solid line above $1~\GeV^2$ is from a NLO 
Altarelli-Parisi fit, while the lines below $1~\GeV^2$ are from   
Regge fits.
\smallskip}
\vskip 1cm
\endinsert

At first it was believed that $\lambda (Q^2)$ could be calculated from
the BFKL equation \ref\fr{J Forshaw and D A Ross,
{\it Quantum chromodynamics and the pomeron
}, Cambridge University Press (1997), and references therein}.
However it was soon realised that this approach could not explain 
the observed rise of $\lambda$ with $Q^2$, nor the large scaling 
violations. Instead, the experimental data are in good agreement \ref\DAS{
R D~Ball and S~Forte, \PL\vyp{B335}{1994}{77}; \vyp{B336}{1994}{77}} with 
with the double-logarithmic rise 
\eqn\das{
F_2 (x,Q^2) \sim \exp(\sqrt{(48/\beta_0)\ln 1/x \ln\ln Q^2}),
}
predicted long ago \ref\DGPTWZ{
De Rujula \etal, \PR\vyp{D10}{1974}{1649}
}
from the lowest-order Altarelli-Parisi equations \ref\esw{
R K Ellis, W J Stirling and B R Webber, {\it QCD and Collider Physics}
Cambridge University Press (1996) and references therein
}. The data can also be fitted in Regge theory \ref\twopom{
A Donnachie and P V Landshoff, \PL\vyp{B437}{1998}{408}},
by adding the exchange of a `hard pomeron' to that of the soft
pomeron; this achieves an effective power $\lambda (Q^2)$ as 
the result of combining fixed-power terms whose relative weights 
vary with $Q^2$.

In this note we review the present difficulties with the BFKL
equation, the uncertainties related to the resummation of 
small $x$ logarithms in Altarelli-Parisi equations, and discuss 
whether either of these approaches is consistent with Regge 
theory and in particular the assumption that the dominant 
singularities are Regge poles. The central question concerns the
extent to which the behaviour of cross-sections in the small $x$
limit may be calculated from perturbative QCD.
 
These are important issues, as the accuracy of any extractions
of parton distribution functions from HERA data and thus of many of 
the predictions for the LHC relies crucially on our understanding of
them. Most of these analyses are currently based on conventional 
fixed order perturbation theory.
                                                          
\topinsert
\centerline{\epsfxsize=0.6\hsize\epsfbox{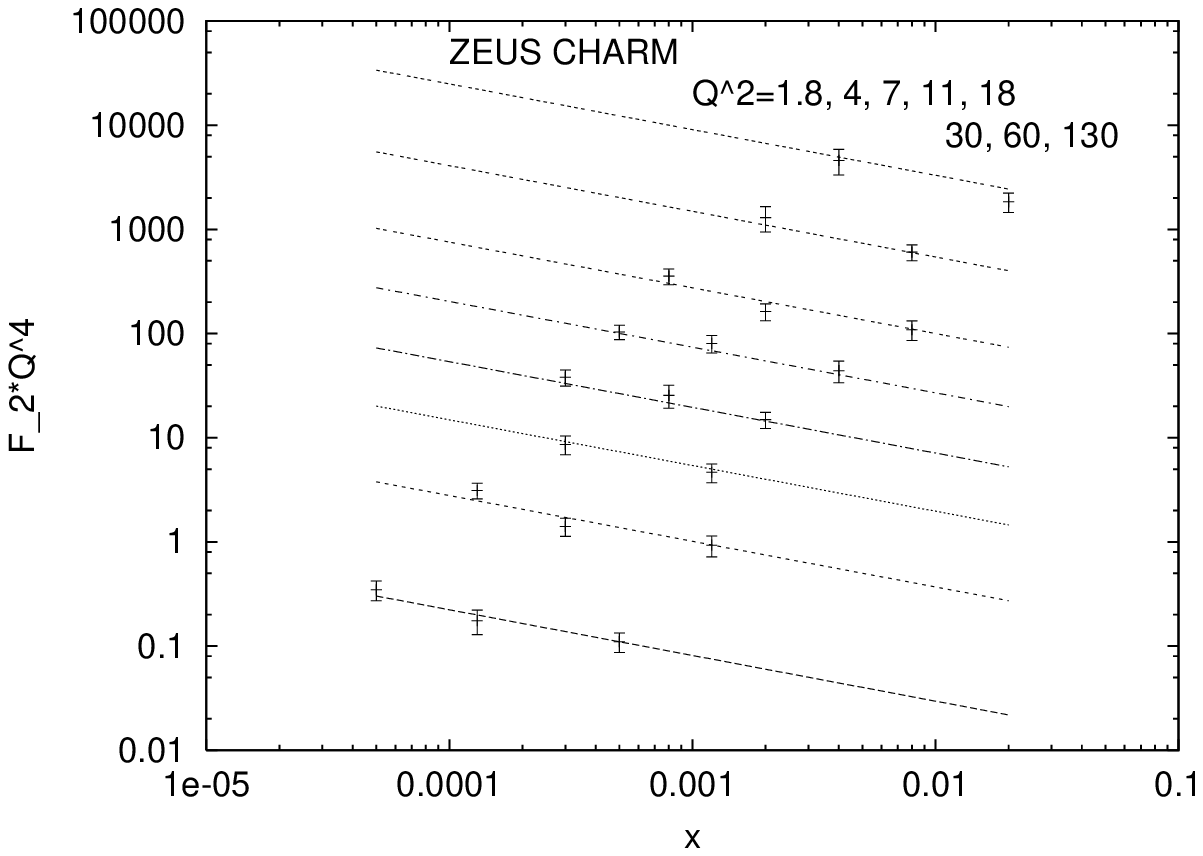}}\semi
\vskip -5mm
\centerline{Figure 2: ZEUS data for $Q^4F_2^c$, fitted \ref\charm{
A Donnachie and 
P V Landshoff, {\tt hep-ph/9910262}}
to a single fixed power of $x$.} 
\endinsert
\bigskip
{\bf The Regge Approach}

The ZEUS collaboration has recently published \ref\zeuscharm{
ZEUS collaboration: A Breitweg et al, hep-ex/9908012}
new data on events in
which a $D^*$ particle is produced, which they use to extract the
contribution $F^c_2 (x, Q^2 )$ to the complete structure function
$F_2 (x, Q^2 )$ from events where the $\gamma^*$ is absorbed by a
charmed quark.  Their data for $F_2^c (x, Q^2  )$ have the 
property \charm\ that, over a wide range of $Q^2$ they can be described 
by a fixed power of $x$:
\eqn\charm{
F_2^c (x, Q^2 ) = f_c (Q^2 ) x^{-\epsilon_0}
}
with $\epsilon_0 \approx 0.4$ and $f_c(Q^2)$ fitted to the data: see
figure~2.

If the behaviour \charm~were literally true, it would imply 
that the Mellin transform $F_2^c(j, Q^2 )$ would have a
pole at $j = 1+\epsilon_0$. Such poles in the complex angular momentum
plane are called Regge poles, and the theory of Regge poles has a long
history \ref\collins{
P D B Collins, {\it Introduction to Regge theory}, Cambridge University
Press (1977)}. 
It has been used very successfully to correlate together a huge amount
of data from soft hadronic reactions: total cross-sections such as
$pp$ and  $\bar{pp}$, partial cross-sections such as $\gamma p \rightarrow
\rho p$, differential cross-sections such as $pp
\rightarrow pp$, and diffraction dissociation (events where the
final state has a very fast hadron).   It is well
established \sigtot\ that ${j}$-plane amplitudes have a pole near to 
$j =\half$, resulting from vector and tensor
meson exchange, and another singularity, 
called the soft-pomeron
singularity, near to $j=1$. It is possible to obtain a good description
of the soft hadronic data by assuming that this singularity too 
is a pole, at $j=1.08$.
Its dynamical origin is poorly understood \ref\softpom{
P V Landshoff and O Nachtmann, \ZP\vyp{C35}{1987}{405}\semi
O Nachtmann, \AP\vyp{209}{1991}{436}\semi
H G Dosch, E Ferreira and A Kramer, \PR\vyp{D5}{1992}{1994}  
}; 
it is presumably the result of some kind of nonperturbative 
gluonic exchange, or perhaps glueball exchange.

While the assumption that the soft-pomeron singularity is a pole
describes a large amount of data well, 
Regge theory admits other types of singularity. 
For example, powers of logarithms of $W^2$ have been used 
to obtain equally good fits to total-cross-section data \ref\softlogs{
P Desgrolard \etal,  \PL\vyp{B309}{1993}{191};\vyp{B459}{1999}{265}\semi
J R~Cudell \etal, {\tt hep-ph/9908218}}. These fits have the 
advantage that they automatically satisfy standard unitarity bounds when
extrapolated to arbitrarily high $W^2$, but they have the disadvantage
that Regge factorization and quark counting rules become rather harder to
understand. Nor can they readily be extended to other applications, such
as \ref\elastic{
A Donnachie and P V Landshoff, \NP\vyp{B267}{1986}{690}} 
$pp$ and $\bar pp$ elastic scattering, and diffraction dissociation
\ref\hone{H1 collaboration: C. Adloff et al, \ZP\vyp{C74}{1997}{221}}. 

Regge theory should be applicable whenever $W^2$ is much greater than all
the other variables, in particular when $W^2\gg Q^2$ (and thus  
$x\ll  1$), even if $Q^2$ is large. However, the tensor-meson
and soft-pomeron poles are insufficient to fit all the HERA
$F_2$ data. An excellent fit can be obtained \twopom\ 
by including a further fixed pole at $j=1+\epsilon_0$, so that
\eqn\hardpom{
F_2 (x,Q^2 ) = \sum_{i=0,1,2} f_i (Q^2 ) x^{-\epsilon_i}
}
This ansatz fits the data all the way from photoproduction 
at $Q^2 = 0$ to $Q^2 = 2000$ GeV$^2$, the highest value 
available at small~$x$.  The soft-pomeron power
is $\epsilon_1=0.08$, the tensor-meson power is $\epsilon_2\approx -0.5$,
while the new power is $\epsilon_0 \approx 0.4$, which we have
already seen is what is needed to fit the data for $F_2^c$ shown in figure~2.
The new leading singularity at
$j=1+\epsilon_0$ is sometimes referred to as the `hard pomeron'
singularity. This does not explain what causes it: it has often been
conjectured that its origin is perturbative QCD, and we will see below
the extent to which it is consistent with our current understanding
based on the summation and resummation of small $x$ logarithms.

Although there is no sign of any contribution from the hard pomeron
in data for purely hadronic processes, it does seem to be present in
$F_2 (x,Q^2 )$ even at extremely small $Q^2$:
measurements \ref\zeus{
ZEUS collaboration: talk by C Amelung at DIS99, Zeuthen}
indicate that even for $Q^2$ as low as 0.045 GeV$^2$, $F_2$ is
rising quite steeply in $x$. Even at $Q^2=0$ the
effective power $\lambda$ may well be greater than that associated
with soft purely-hadronic collisions.

Similarly\twopom,
the data for $\gamma p \rightarrow J/\psi\, p$ are 
described well by the sum of two powers in the amplitude, 
$(W^2 )^{\epsilon_0}$ and 
$(W^2 )^{\epsilon_1}$ at $t=0$.  One does not expect a contribution 
from tensor meson exchange, because of Zweig's
rule. The Regge picture also successfully describes the differential
cross-section away from $t=0$. 

The striking feature of these fits is that such a wide variety of 
different data may be described using a simple parameterization: this
suggests a universal underlying mechanism, and raises the hope that
the hard component at least might be derivable from perturbative QCD.
However, the $j$- plane singularities need not be poles, 
so the $x$ dependence need not
be simple powers of $x$: powers of $\ln 1/x$ could do as well. 
Furthermore, Regge theory does not determine the 
coefficient functions $f_i (Q^2 )$ in {\hardpom}. 
Nor is it clear that three terms in \hardpom\ will
always be enough: as the range in $x$ and $Q^2$ increases still further, it
may be that yet more terms are required.

Thus although the $x$ and $Q^2$ of the existing data can be fitted
using a Regge pole ansatz, the uncertainties in any extrapolation
outside the existing kinematic range (such as from HERA to the
LHC) are difficult to quantify. 
Moreover, it is not possible using Regge theory alone to predict
jet cross sections, or indeed vector boson or top or Higgs production 
cross sections: we need more dynamics. Our only candidate for a
complete theory of strong interactions at high energies is 
perturbative QCD, and it is to the understanding
of perturbative QCD at small $x$ that we now turn.

\bigskip\noindent
{\bf  QCD: Resummation of Logs of \pmb{$x$}}
\smallbreak

At first it was hoped that the BFKL equation provided a purely
perturbative calculation of the value of $\lambda
(Q^2)$.  This hope was based on the leading contribution to the BFKL
kernel $K(Q^2,k^2)$ with fixed coupling. Its Mellin transform
$\chi(M)$ has a minimum at $M=\half$, which gives rise to a 
power rise of the form $x^{-\lambda}$, with 
$\lambda=\lambda_0\equiv\chi(\half)=12\ln2\alpha_s/\pi$,
in qualitative agreement with the first data sets. However 
this agreement was superficial, essentially
because the $Q^2$ dependence was incorrect (see figure~1): $\lambda$ did not
rise with $Q^2$, but remained fixed. There were suggestions that 
this was because the BFKL equation did not take sufficient account of
energy conservation and of nonperturbative effects \ref\bfklproblems{
J C Collins and P V Landshoff, \PL\vyp{B276}{1992}{196}\semi
M F  McDermott, J R  Forshaw and G G  Ross, 
\PL\vyp{B349}{1995}{189}\semi
J Bartels, H Lotter and M Vogt, \PL\vyp{B373}{1996}{215}}: it 
is difficult to avoid important contributions from soft gluons, which 
cannot be estimated using perturbation theory. 
For this reason attempts to improve the kernel by making the coupling run were
never entirely successful \ref\morebfklproblems{
L P A Haakman \etal, \NP\vyp{B518}{1998}{275}\semi
Y V Kovchegov and A H Mueller \PL\vyp{B439}{1998}{428}\semi
N Armesto \etal, \PL\vyp{B442}{1998}{459}
}: running couplings make the equation unstable, leading to
unphysical effects. 

The full extent of the difficulties was reinforced by
the calculation of the next-to-leading order correction to the 
kernel \ref\fl{
V~S~Fadin and L~N~Lipatov, \PL\vyp{B429}{1998}{127}}: 
the correction turned out to be very large and negative,
inverting the minimum of the BFKL function $\chi(M)$, which 
was responsible for 
the power behaviour at leading order (see figure~4a). Since the saddle
points of the inverse Mellin transform were now off the real axis, the
NLLx equation gave rise to negative cross-sections in the 
Regge region \ref\ross{D A Ross, \PL\vyp{B431}{1998}{161}}. 
This destroyed any faith that might have remained in the 
leading-order prediction.

Various proposals to fix up the BFKL equation have been put forward: 
for example a particular choice of the renormalization 
scale \ref\bfklp{S J~Brodsky 
\etal,  {\it JETP~Lett.~}\vyp{70}{1999}{155}\semi
R S~Thorne, \PR\vyp{D60}{1999}{054031}}, or a different 
identification of the large logs which are resummed \ref\schmidt{
C R~Schmidt, \PR\vyp{D60}{1999}{074003}}.
However the root of the problem \ref\salam{G~Salam, \JHEP\vyp{9807}{1998}{19}} 
is that the perturbative contributions to $\chi(M)$
become progressively more and more singular at integer values of $M$,
due to unresummed logarithms of $Q^2$ and $k^2$ in the kernel $K$. 
In particular, near $M=0$ the expansion oscillates wildly.
It follows that a perturbative expansion which sums 
logarithms of $x$ must also resum the large logarithms of $Q^2$ to all
orders in perturbation theory if it is to be useful.

\bigskip\noindent
{\bf QCD: Resummation of Logs of $Q^2$}

The usual way to resum logarithms of $Q^2$ is to use Altarelli-Parisi
evolution equations, with the splitting functions calculated at a
given fixed order in perturbation theory. If one starts at some 
initial scale $Q^2_0$ with parton distributions that rise less steeply than  
a power in $1/x$, then fixed order evolution to
higher $Q^2$ leads to distributions that become progressively 
steeper in $1/x$ as $Q^2$ increases, in agreement with 
the $F_2$ data from HERA. More significantly the prediction\DGPTWZ\
of the specific form \das\ of the rise is in good agreement \DAS\ with the data
over a wide region of $x$ and $Q^2$.
This is widely seen as a major triumph for perturbative QCD, as 
direct evidence for asymptotic freedom \ref\af{
F Wilczek, Dirac medal lecture, Trieste, 1994 {\tt hep-th/9609099}}: 
the coefficient $\beta_0$ in \das\ which
determines the slope of the rise is the first coefficient of
the QCD $\beta$-function. 

\topinsert
\centerline{\epsfxsize=0.4\hsize\epsfbox{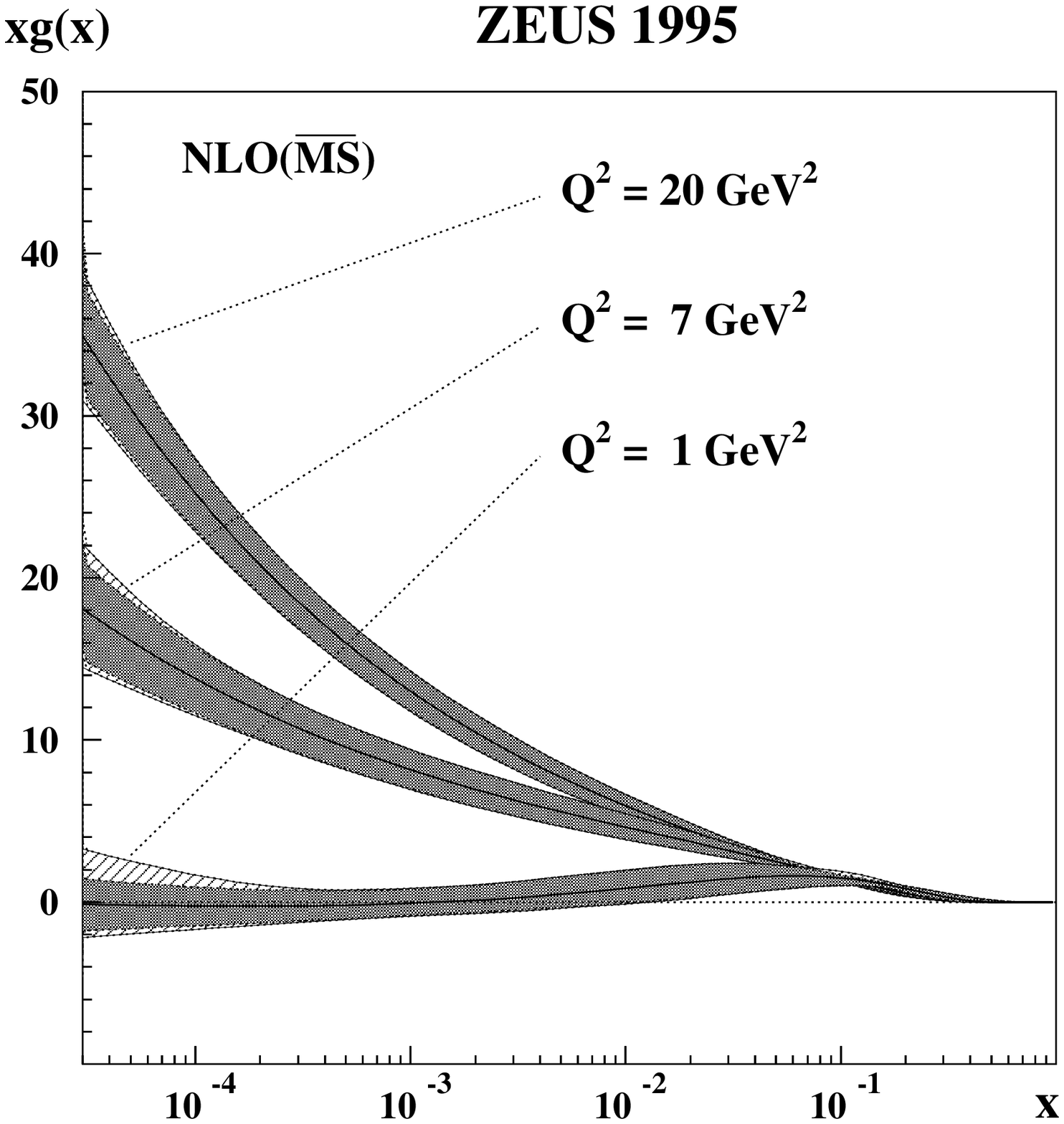}}\semi
\centerline{\epsfxsize=0.5\hsize\epsfbox{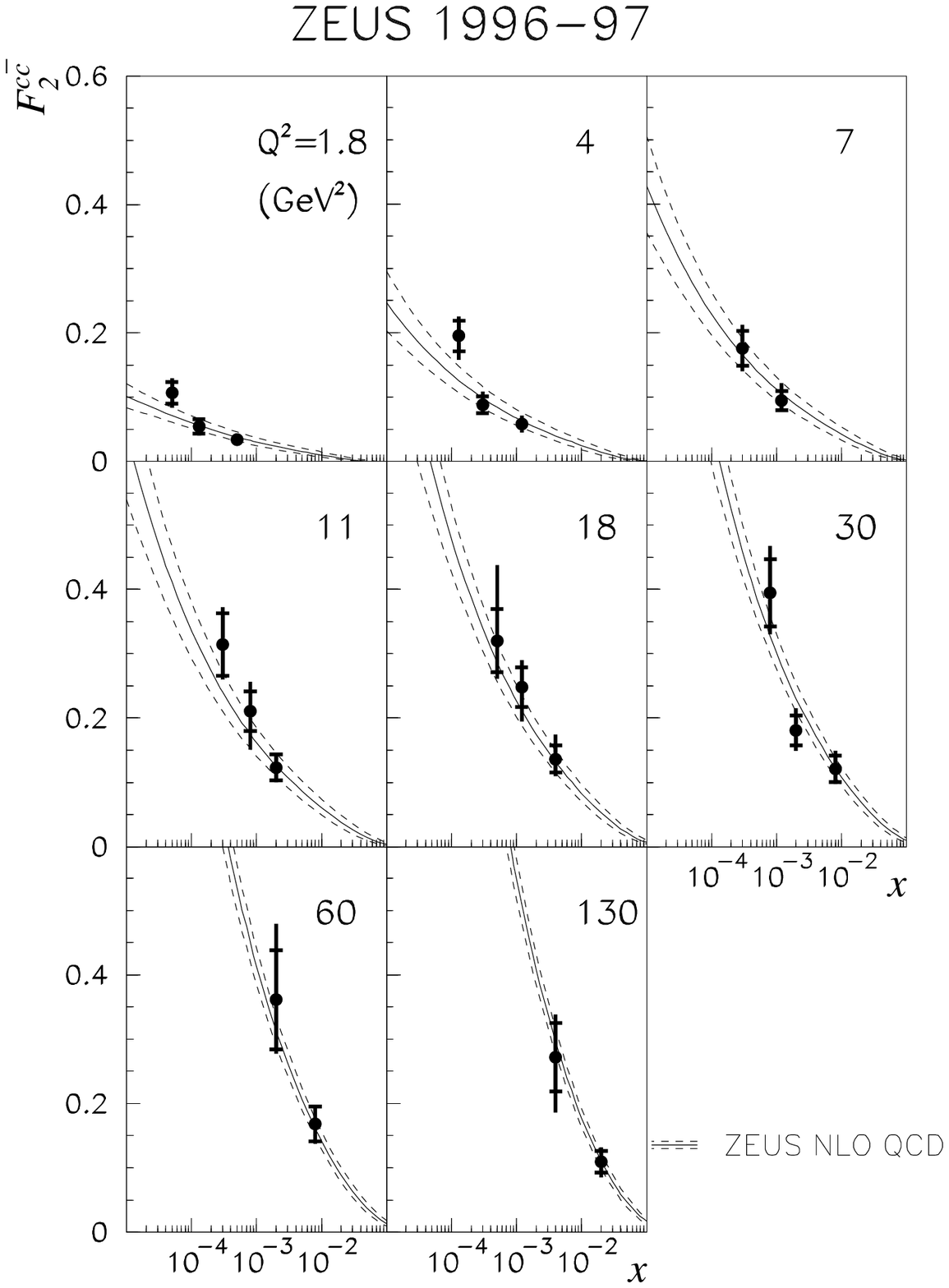}}\semi
{\parindent=1cm \narrower\noindent
Figure 3: a) The gluon
distribution extracted from a NLO fit to ZEUS data for $F_2$
\ZEUS. 
b) The ZEUS data for $F_2^c$ \zeuscharm, compared
to the QCD prediction obtained from the gluon a).
\vskip 2cm}
\endinsert

The success of fixed-order perturbative QCD
in describing the increasingly precise HERA $F_2$ data 
when $Q^2\gsim 1~\GeV^2$ has been confirmed many times 
by successful NLO fits \ref\fits{
See for example M~Botje (ZEUS Collaboration ), {\tt hep-ph/9905518}\semi
V~Barone, C~Pascaud and F~Zomer, {\tt hep-ph/9907512}}.   
From these a gluon distribution may be extracted, (see figure 3a),
and predictions for $F_2^c$ (figure~3b), dijet production, 
and $F_L$, all of which
have now been supported by direct measurements \ref\HERArevs{
See {\it e.g.} M~Klein, Lepton-Photon proceedings (Stanford, 1999)\semi 
P~Marage, ICHEP proceedings (Tampere, 1999), {\tt hep-ph/9911426}}.
Clearly fixed order perturbative QCD works well at HERA: none of these
predictions is trivial, and all are successful. 
Of course once  $Q^2_0$ is as small as $1~{\rm GeV}^2$ or less 
a perturbative treatment is no longer appropriate, and
indeed an instability develops in the NLO gluon distribution at around
such a scale (see figure 3a).

It is perhaps useful to compare figure~2 with figure~3b: the 
data are the same on each figure, but the curves
on the former are the result of a power fit that assumes a flavour-blind
hard pomeron,
while those on the latter are from a
straightforward parameter-free prediction made using NLO perturbative 
QCD. Interestingly the conclusions are also different: the slope of
the rise in $x$ manifestly increases with $Q^2$ in figure 3b
(corresponding to the rise of the slopes in figure 1a and figure 3a), 
while in figure 2 it is fixed.

It is important to realise that the success of the NLO perturbative QCD
predictions is crucially dependent on the nonperturbative input at 
the initial scale $Q^2_0\sim 1 \GeV^2$ being `soft' --- not rising too quickly 
with $x$ --- so that the rise in $x$ can be generated dynamically. If
instead the rise were input in the form \hardpom, growing as $x^{-\epsilon_0}$
with $\epsilon_0$ as large as $0.4$, this would when evolved perturbatively
with the NLO anomalous dimension
lead to a $Q^2$ dependence which was independent of $x$ and thus 
inconsistent with the data \DAS\ (see figure~1). If one were to 
insist on such a hard pomeron singularity, one would thus to be 
consistent also have to argue that NLO perturbative QCD could not be
applied in this region. The many quantitative successes of NLO 
perturbative QCD at HERA \refs{\DAS,\fits,\HERArevs} would then 
have to be considered merely fortuitous. Conversely, if one instead 
accepts that the success
of the perturbative predictions is significant, one would then have 
to conclude that the simple assumption \hardpom\ that the 
rightmost singularity in the $j$-plane is a simple pole is incorrect,
since the perturbative results rely for their success 
on a soft input.

This said, to obtain reliable predictions for processes at the LHC it is
not sufficient to confirm NLO QCD within experimental errors at 
HERA: we must also be able to understand theoretical errors.
In particular, at small $x$ the approximation to the splitting
functions given by retaining only the first few terms in an expansion 
in powers of $\alpha_s$ is not necessarily very good: 
as soon as $\xi=\log{1/x}$ is sufficiently large that $\alpha_s \xi\sim 1$, 
all terms of order $\alpha_s (\alpha_s \xi)^n$ (LLx) and 
$\alpha_s^2 (\alpha_s \xi)^n$ (NLLx) must also be considered in order 
to achieve a result which is reliable up to terms of order $\alpha_s^3$.
In fact $\alpha_s\xi\gsim 1$ throughout most of the HERA kinematic 
region, so one might expect these effects to be significant.
The fact that empirically they seem to be small is thus a mystery 
requiring some explanation.

This argument may be sharpened by consideration of the $j$-plane
singularities of the Mellin transform $F_2(j, Q^2 )$. At the $n$-th order 
in fixed order perturbation theory the iteration 
of small $x$ logarithms in the evolution gives rise 
to essential singularities of the form 
\eqn\ess{
(j-1)^{-1}\exp (\alpha_s^n/(j-1)^n)
}
The $j=1$ singularity thus becomes more severe order by order in 
perturbation theory. This is not necessarily a problem phenomenologically, 
since \ess\ corresponds to a sequence of predictions for 
measurable quantities such as $F_2(x,Q^2)$ that are  strictly 
convergent \ref\sum{R D Ball and S Forte, \PL\vyp{B351}{1995}{313}} 
provided only that $x>0$. It follows that although \ess\ may 
not be correct actually at the point $j=1$ it may be a good numerical 
approximation to the correct behaviour away from $j=1$.

Furthermore there is good reason to believe
that a resummation over all orders $n$ might remove the singularity 
\ref\cdl{
J R Cudell, A Donnachie and P V Landshoff, \PL\vyp{B448}{1999}{281}
}. The argument is that, if there is a singularity
at a fixed point in the complex $j$-plane for large values of $Q^2$, such
as a naive application of \ess\ might seem to imply, then 
considerations of analyticity in $Q^2$ suggest that it might 
also be present at small $Q^2$.
While this is not completely excluded, the Mellin transform 
variable $j$ is essentially a complex angular momentum and studies made
more than a quarter of  a century ago \ref\fixed{
P V Landshoff and J C Polkinghorne,
\PR\vyp{D5}{1972}{2056}} never found any need for a worse singularity 
than a fixed pole at $j=1$ in Compton-scattering amplitudes, 
with no singularity at all at that point in $F_2$. 

The problem with this argument is that although it suggests that the
singularity structure \ess\ is incorrect, it still doesn't tell us 
precisely what or where the rightmost singularities are in the
$j$-plane. Furthermore it is clearly not possible 
to deduce precisely what it is from the data: to do this we 
would need to do experiments of arbitrarily high precision at 
arbitrarily high energies. It is 
thus interesting to ask whether we can instead deduce it from 
perturbative QCD. To do this, we would at least need a sensible resummation 
of small $x$ logarithms. We now discuss the difficult problem of 
constructing such a resummation. 

\topinsert
\centerline{\epsfxsize=0.6\hsize\epsfbox{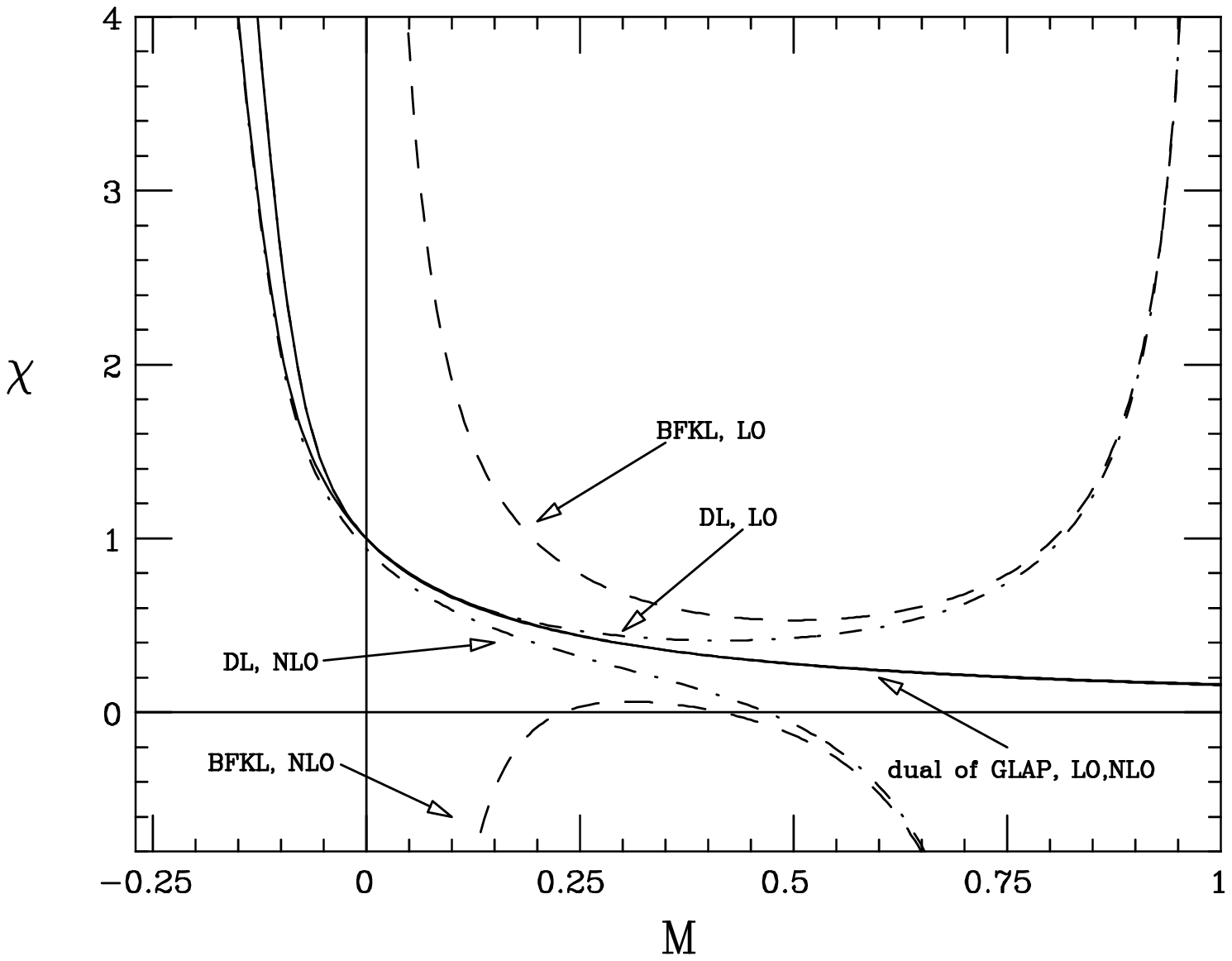}}\semi
\centerline{\epsfxsize=0.6\hsize\epsfbox{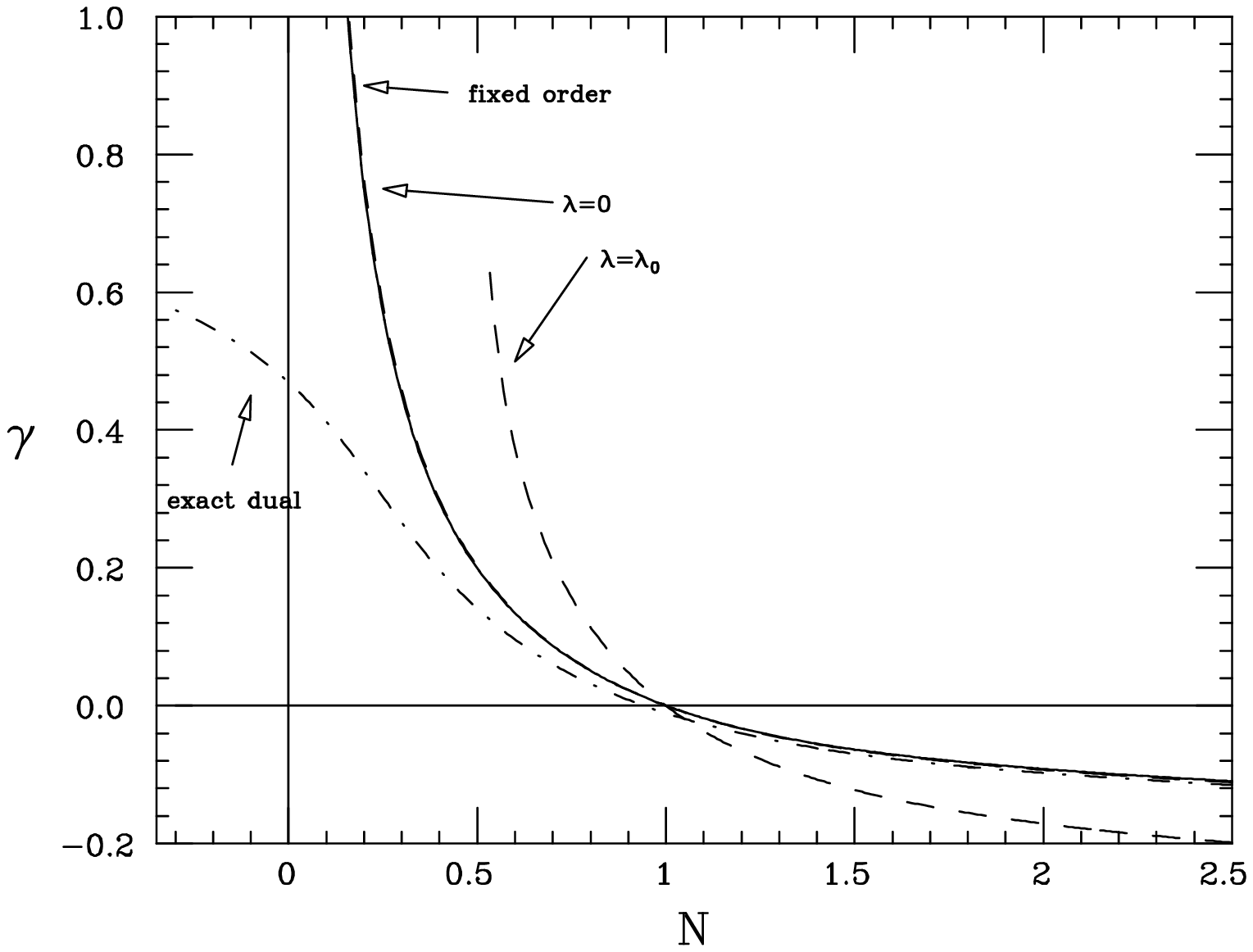}}
\medskip
{\parindent=1cm \narrower\noindent
Figure 4: (a) the BFKL function $\chi(M)$ and (b) the corresponding anomalous
dimension $\gamma(N)$ in various approximation schemes
\ref\ABF{G~Altarelli, R~D~Ball and S~Forte, {\tt hep-ph/9911273}}.
\smallskip}
\endinsert
 
\bigskip\noindent
{\bf QCD: Resummation of Logs of $x$ and Logs of $Q^2$} 

\nref\Jaro{T~Jaroszewicz, \PL\vyp{B116}{1982}{291}}
\nref\sumtoo{M~Ciafaloni, \PL\vyp{B356}{1995}{74}\semi
R D Ball and S Forte, \PL\vyp{B359}{1995}{362}\semi
S~Catani, \ZP\vyp{C70}{1996}{263}; \ZP\vyp{C75}{1997}{665}\semi
G~Camici and M~Ciafaloni, \NP\vyp{B496}{1997}{305}}
\nref\res{R D Ball and S Forte, \PL\vyp{B465}{1999}{271}}
Using the BFKL kernel it is possible \Jaro\ 
to deduce the coefficients of the LLx 
singularities of the splitting function to all orders in perturbation 
theory, ie of all terms in the anomalous dimension $\gamma(N)$ of
the form $\alpha_s^n/N^n$, where $N=j-1$. Summing up these 
singularities converts the sum of poles into a cut starting from 
$N=\lambda_0$, apparently confirming the Regge expectation about 
the behaviour at $j=1$: it is this cut which at fixed coupling gives the 
power rise of the BFKL pomeron. This procedure may be extended beyond
LLx \refs{\sum,\sumtoo,\res}: the anomalous dimension
$\gamma(\alpha_s,N)$ in a particular factorization scheme (such as
\MS) is related to a BFKL function $\chi(\alpha_s,M)$ through the `duality'
relation 
\eqn\duality{\chi(\alpha_s,\gamma(\alpha_s,N))=1.}
Expanding this relation to NLLx, and using calculations of 
the coefficient function and
gluon normalization \ref\CH{S~Catani and F~Hautmann, 
\PL\vyp{B315}{1993}{157}; \NP\vyp{B427}{1994}{475}} and of 
the NLLx kernel \fl, we can compute the coefficients of all
terms of the form  $\alpha_s\alpha_s^n/N^n$ in the anomalous
dimension. Such an approach has several advantages over the direct solution of
the BFKL equation: there is a clean factorization of hard and soft 
processes, running coupling effects are properly taken care of
by well formulated renormalization group arguments, and it is easy 
to arrange for a smooth matching to the large $x$ region. 

\nref\mad{R~D~Ball and S~Forte, {\tt hep-ph/9805315}\semi
J Bl\"umlein et al., {\tt hep-ph/9806368}} 
However it was known some time ago that reconciling the summed 
logarithms with the HERA data was actually rather difficult \ref\sumfits{
R K~Ellis, F~Hautmann and B~R~Webber, \PL\vyp{B348}{1995}{582}\semi
R D Ball and S Forte \PL\vyp{B358}{1995}{365} and {\tt hep-ph/9607291}\semi
I~Bojak and M~Ernst, \NP\vyp{B508}{1997}{731}}.
Once all the NLLx corrections were known it became clearer why: 
the expansion in summed anomalous dimensions at LLx, NLLx,\dots is unstable 
\refs{\res,\mad}, the ratio
of NLLx/LLx contributions growing rapidly as 
$\xi=\log{1/x}\to\infty$. 
It follows that the previous theoretical estimates 
of the size of the effects of the small $x$ logarithms 
based on the fixed order BFKL equation, either at LLx or NLLx, 
were all hopelessly unreliable. Indeed any calculation 
which resums LO and NLO logs of $Q^2$, but sums up only 
LO and NLO logarithms of $x$ is seen to be 
insufficient: some sort of all order resummation of the small 
$x$ logarithms is always necessary. Clearly there are many ways in which such
a resummation might be attempted: what is needed are guiding
principles to keep it under control.

One such principle is momentum conservation \ABF: before using $\chi(M)$ to
compute the corrections to $\gamma(N)$ through the duality
eqn.\duality, we should first resum all the
LO and NLO singularities at $M=0$ discussed above, and impose the momentum
conservation condition $\gamma(\alpha_s,1)=0$, whence (from eqn.\duality) 
$\chi(\alpha_s,0)=1$. Since these are collinear
singularities, their coefficients may be determined from the usual LO and
NLO anomalous dimensions, again using the duality relation
eqn.\duality, but this time in the reverse direction. It turns out 
that when the $M=0$ singularities are resummed they account for almost
all of $\chi$ in the region of $M=0$ (see figure 4a): this explains 
already why the remaining small $x$ corrections have not yet been seen
at HERA. Small $x$ logarithms are simply numerically much less important than
collinear logarithms.

The second principle is perturbative stability. The instability found
at NLLx can be shown to follow inevitably from the shift in 
the value $\lambda$ of $\chi$ at the minimum due to
subleading corrections \res. This shifts the position of the  
singularity from $N=\lambda_0$ to $N=\lambda_0+\Delta\lambda$, and
this shift must be accounted for exactly if a sensible resummed 
perturbative expansion is to be obtained. Since in practice 
the correction $\Delta\lambda$ is of the same order as the leading
term $\lambda_0$, it seems probable that 
$\lambda=\lambda_0+\Delta\lambda$ is not calculable in perturbation
theory: rather the value of $\lambda$ may be used to parameterise the
uncertainty in the value of $\chi$ in the vicinity of $M=\half$.

This uncertainty is clearly due to the unresummed infrared logarithms
at $M=1$. In \ref\ciaf{
M~Ciafaloni \etal, \PL\vyp{B452}{1999}{372};
\PR\vyp{D60}{1999}{114036}; \JHEP\vyp{9910}{1999}{017}} an attempt is
made to resum these singularities through a symmetrization of $\chi$ 
about $M=\half$:  $\chi$ is then supposedly determined for all 
$0\le M\le 1$, and $\lambda$ is given by the height of its minimum.   
The main shortcoming of this approach is that it makes implicit 
assumptions about the validity of perturbation 
theory when $Q^2$ is very small.

Putting together the two principles of momentum conservation and
perturbative stability, we can compute fully resummed NLO anomalous
dimensions (see figure 4b). The result depends on the unknown parameter
$\lambda$. Provided $\lambda\lsim 0$, the corrections to
Altarelli-Parisi evolution in the HERA region are tiny: for larger
values they may be significant at low $x$ and low $Q^2$, and it might
then be possible to determine $\lambda$ from the data. It can be
seen from the plot that the singularity structure at $N=0$ (and thus $j=1$)
is still completely undetermined: this is a reflection of the  
uncertainty in the $\chi$ plot at $M=1$, which makes it not only
unclear as to the value of $\chi$ at its minimum, but even whether
there is a minimum at all. To determine the position and nature of the
rightmost singularities in the $j$-plane would presumably require control of
$\chi(M)$ at $M=1,2,\ldots$, which is clearly beyond current 
perturbative technology.

It seems that to make further progress we require either genuine
nonperturbative input, or a substantial extension of the perturbative
domain. A possible way in which this might be done through a 
new factorization procedure was explored in \ref\afp{R D Ball and
S Forte, \PL\vyp{B405}{1997}{317}}, from which the main conclusion
was that at small $x$ the coupling should run not with $Q^2$, but with
$W^2$. Preliminary calculations \ref\rgr{
R G Roberts, \EPJ\vyp{C10}{1999}{697}}  suggest that this is not
phenomenologically unnacceptable. However much more work remains to be
done.

\bigskip\noindent
{\bf Summary}

At low $Q^2$ but high $W^2$ Regge theory works well and gives
nontrivial and successful predictions. At high $Q^2$ and 
small $x$ NLO perturbative QCD works well and gives nontrivial
and successful predictions, with quantifiable uncertainties 
due to the need for a controlled resummation of small $x$ logarithms. 
In the same region, Regge theory can also fit data successfully, 
but without the predictive power of perturbative QCD. Neither 
Regge theory, nor conventional perturbative QCD, nor even the 
data, seem to be able to predict the precise form 
of cross sections in the Regge limit $W^2\to\infty$ with 
$Q^2$ large. To do this, new ideas will probably be needed.

\bigskip\noindent
{\bf Acknowledgements:} RDB would like to thank Guido~Altarelli,
Stefano~Catani, John~Collins, Gavin~Salam, Dave~Soper, 
and Andreas Vogt for discussions on this subject, and in particular
Stefano~Forte for a critical reading of the manuscript.

\bigskip
\footatend\vfill\immediate\closeout\rfile\writestoppt
\baselineskip=11pt\parskip=3pt
\centerline{{\bf References}}\bigskip{\frenchspacing%
\parindent=20pt\escapechar=` \input refs.tmp\vfill}\nonfrenchspacing
\bigskip
\vfill
\font\eightit=cmti8
\bigskip{\eightit
This research is supported in part by the EU Programme
``Training and Mobility of Researchers", Networks
``Hadronic Physics with High Energy Electromagnetic Probes"
(contract FMRX-CT96-0008) and
``Quantum Chromodynamics and the Deep Structure of
Elementary Particles'' (contract FMRX-CT98-0194),
and by PPARC}
\vfill

\bye